\documentclass{aastex}
 
\shorttitle{CO (7$\rightarrow$6) and (4$\rightarrow$3) Observations near the 
Galactic Center}
\shortauthors{Kim et al.}

\newcommand \kms  {km~s$^{-1}$}

\slugcomment{}
\begin{document}
 
\title{AST/RO Observations of CO $J = 7 \rightarrow 6$ and $J = 4 \rightarrow 3$
Emission toward the Galactic Center Region}

\author{Sunguen Kim\altaffilmark{1}, Christopher L. Martin\altaffilmark{2},
Antony A. Stark\altaffilmark{3}, Adair P. Lane\altaffilmark{4}}

\affil{Harvard-Smithsonian Center for Astrophysics}
\affil{60 Garden Street, MS-12}
\affil{Cambridge, MA 02138}

\altaffiltext{1}{skim@cfa.harvard.edu}
\altaffiltext{2}{cmartin@cfa.harvard.edu}
\altaffiltext{3}{aas@cfa.harvard.edu}
\altaffiltext{4}{adair@cfa.harvard.edu}

\begin{abstract}

We present position-velocity strip maps of the Galactic Center region 
in the CO $J=7 \rightarrow 6$ and $J=4 \rightarrow 3$ transitions 
observed with the Antarctic Submillimeter Telescope and Remote 
Observatory (AST/RO) located at Amundsen-Scott South Pole Station.
Emission from the two rotational transitions of $^{12}$CO was mapped at
$b=0^{\circ}$ for $3.5^{\circ}> \ell > -1.5^{\circ}$, on a $1'$ grid 
with a FWHM beamsize of $58''$ at 806 GHz and $105''$ at 461 GHz. 
Previous observations of CO $J=4\rightarrow 3$ (Martin et al., in 
preparation) and [C\,I] (Ojha et al. 2001) emission from this region 
show that these lines are distributed in a manner similar to CO 
$J=1 \rightarrow 0$ (Stark et al. 1987); the (CO 
$J=4 \rightarrow 3$)/(CO $J=1 \rightarrow 0$) line ratio map is
almost featureless across the entire Galactic Center region.
In contrast, the CO $J=7 \rightarrow 6$ emission from the Galactic Center  
is strongly peaked toward the Sgr~A and Sgr~B molecular complexes. 
A Large Velocity Gradient (LVG) analysis 
shows that aside from the two special regions Sgr~A and Sgr~B,
the photon-dominated regions within a few hundred parsecs of the Galactic Center 
are remarkably uniform in mean density and kinetic temperature at $n = 2500$ to
$4000 \, \mathrm{cm^{-3}}$ and T = 30 to 45 K.
The (CO $J=7 \rightarrow 6$)/(CO $J=4 \rightarrow 3$) line 
temperature ratios near Sgr~B are a factor of two higher than those 
observed in the nuclear region of the starburst galaxy M82 (Mao et al. 2000), 
while the CO($J=7 \rightarrow 6$)/CO($J=4 \rightarrow 3$) line temperature 
ratios around Sgr~A are similar to M82.
The line ratio on large scales from the Galactic Center region is an order
of magnitude less than that from M82.
\end{abstract}


\section{Introduction}
\label{s:intro}

Observations of photons emitted by the various rotational transitions 
of the ground vibrational state of carbon monoxide (CO)
are the primary means of studying molecular gas in the Galaxy.
These spectral lines occur at frequencies of 
$J \times 115 \, {\mathrm{GHz}}$,
for the transition from the $J$ to the $J-1$ rotational state.
Numerous galactic surveys (Combes 1991) have studied the 
lowest-frequency $J = 1 \rightarrow 0$ line.
The brightness of this spectral line is roughly proportional to
the total molecular column density within the telescope beam
(Liszt 1984), provided that
the molecular gas has a relatively low 
density ($n < 3 \times 10^3 \,{\mathrm{cm}^{-3}}$) and column 
density ($N < 10^{23} \, {\mathrm{cm}^{-2}}$).  Surveys of the
$J = 1 \rightarrow 0$ line therefore give an indication of the 
extent and distribution of molecular material.

A more complete picture of the thermodynamic state of the molecular 
gas can be gained through observations of other, higher-$J$ 
transitions of CO and the various rotational transitions of the 
isotopically-substituted species $^{13}$CO and C$^{18}$O.  
If the brightness of several of these lines is known, models of the
excitation and radiative transfer can be used to solve for 
the density, temperature, and cooling rate of the molecular gas.
To avoid degeneracy in this solution, it is valuable to have
observations
of a line transition from a $J$-state that is sufficiently high
in energy such that it is only weakly populated. 
CO in interstellar molecular gas is typically in a thermodynamic
state where the low-$J$ transitions are in approximate thermal
equilibrium, so that
all of the rotational transitions below some $J$-level have roughly
the same excitation temperature as the $J = 1 \rightarrow 0$
transition.
The observable quantity that is usually compared with theory 
is the ratio of line brightnesses.
If only low-$J$ lines are observed, it is found that the ratio of
these line brightnesses are only weakly dependent on excitation
temperature and that maps of all the low-$J$ transitions look similar,
regardless of variations in excitation temperature across the map.
If, however, we observe a transition from a high-$J$ level
that is not strongly excited,
it will be found that the line
brightness ratios involving that transition do vary significantly 
across the map, and the value of the
excitation temperature can be more readily determined.

Previous observations with the AST/RO telescope (see Figure 1)
have shown that the distribution of the  CO $J = 4 \rightarrow 3$
line emission is remarkably similar to that of the
CO $J = 1 \rightarrow 0$ line throughout the Galactic Center region.
Previous observations of CO  $J = 7\rightarrow 6$ 
(Harris et al. 1985) showed that line to be unusually bright 
within $\pm 200''$ of the Galactic Center.  In this paper, we report 
observations of the CO $J = 7 \rightarrow 6$ and $J = 4\rightarrow 3$
transitions in a strip 
map several degrees in extent, and find that the $J = 7 
\rightarrow 6$ line has significantly lower brightness temperature 
and significantly different spatial distribution than 
the lower-$J$ CO lines. We can therefore produce a new model of
the thermodynamic state of the CO gas along this 1-dimensional strip.

\section{Observations}
\label{s:obs}

These observations were made from 21 to 25 July 2000 with the Antarctic 
Submillimeter Telescope and Remote Observatory (AST/RO), located at 
Amundsen-Scott South Pole Station. AST/RO is a 1.7~m diameter, offset 
Gregorian telescope capable of observing at wavelengths between 
200~$\mu$m and 1.3~mm (Stark et al. 2001). 

The facility dual-channel SIS waveguide receiver (Walker et al. 1992;
Honingh et al. 1997) was used in position-switching mode.
Emission from the 806.6517~GHz $J = 7 \rightarrow 6$ transition 
of $^{12}$CO was mapped at $b$ = 0$^{\circ}$, 
3.5$^{\circ} > l > -$1.5$^{\circ}$ with 1$'$ 
spacing and a beam size of $58''$.  The 461.041~GHz 
$J = 4 \rightarrow 3$ transition of $^{12}$CO was observed 
simultaneously, with a beam size of $105''$.  The atmosphere-corrected 
system temperature during the observations ranged from 12,000 to 
26,000~K at 806~GHz and from 3000 to 4000~K at 461~GHz.  Observing 
time at each of the 304 observed positions was typically 7 minutes.

The intermediate frequency output of each receiver was connected 
to one of two acousto-optical spectrometers (Schieder, Tolls, and 
Winnewisser 1989) of which $900 \times$ 1-MHz-wide channels were 
in use at 806~GHz and $1000 \times$ 1-MHz-wide channels were in 
use at 461~GHz.  This configuration provided a velocity range of 
225 \kms\ at 806~GHz and 438 \kms\ at 461~GHz.  The channel spacing 
of 0.67~MHz corresponds to a velocity resolution of 0.37 \kms\ at 
806~GHz and 0.65 \kms\ at 461~GHz. 
The high frequency observations were made with the 
CO $J = 7 \rightarrow 6$
line in the lower sideband (LSB). Since the intermediate frequency 
of the AST/RO system is 1.5~GHz, the $^3P_2 \rightarrow ^3P_1$ line 
of [C\,I] at 809.34197~GHz appears in the upper sideband and is 
superposed on the observed LSB spectrum.  The local oscillator 
frequency was chosen so that the nominal line centers appear 
separated by $\sim$100 \kms\ in the double-sideband
spectra.  Velocity increases with spectrometer channel number
in opposite directions for the two sidebands; in Figure 4, for 
example, velocity increases to the left for the [C\,I] line and 
to the right for the CO.  There is 809~GHz [C\,I] emission 
superposed on the 806~GHz CO
$J = 7 \rightarrow 6$ data at most positions where the
$J = 7 \rightarrow 6$ line is strong.
The central channel of the spectrometer was shifted
for observations at negative longitudes so that the limited
bandpass covered most of the velocity range of known low-$J$ 
emission.

During the time these observations were made, the 
AST/RO automated calibration system (Stark et al. 1997, 2001)
was broken, so that we were not able to make automated sky
or receiver measurements.  We did, however, make
manual measurements once a day.  These show that the receiver 
gain and noise were stable within a few percent throughout 
the period of the observations.  To determine the atmospheric
opacity and sky temperature, we used data from the 
NRAO-CMU $350 \, \mu {\mathrm{m}}$ tipper in operation at the 
South Pole.  Several times each hour, this instrument makes 
a calibrated skydip using a broadband room-temperature 
bolometer.  During the subsequent observing season,
when the automated calibration system was working, it was
determined that the opacity in the narrow 806~GHz band
is related to the opacity measured by the NRAO-CMU tipper
through the relation (R. A. Chamberlin, private communication): 
$$\tau_{806 \mathrm{GHz}} =  1.82\  \tau_{\mathrm{NRAO-CMU}} - 1.51 \, .$$
The relation used to calibrate the 461 GHz data was (Chamberlin 2001):
$$\tau_{461 \mathrm{GHz}} =  0.55\  \tau_{\mathrm{NRAO-CMU}} - 0.25 \, .$$
We used these relations, together with the NRAO-CMU tipper data and 
thermometer measurements of the outside ambient temperature, to 
calculate the $T_{\mathrm{sky}}$ value we would have gotten if the 
automated calibration system had been working.  These values were 
edited into a copy of the data file, allowing the data to be treated 
normally by our data reduction program, {\it COMB}, as described in 
Stark et al. (2001).

\section{Results and Analysis}
\label{s:results}
\subsection{Strong CO $J = 7 \rightarrow 6$ Emission in the 
Sgr~B and Sgr~A Complexes}

Longitude-velocity maps of the $^{12}$CO $J = 4 \rightarrow 3$ and 
$J = 7 \rightarrow 6$ distributions are shown in Figure 2.  As can be
seen by reference to the $(\ell,\, b)$ maps in Figure 1, the strip 
maps at $b=0$ shown in Figure 2 skirt the outer regions of the 
strongest CO features, rather than passing through their peak positions.  
Nonetheless, the CO $J = 7 \rightarrow 6$ emission is concentrated in the 
Sgr~B and Sgr~A complexes at velocities between 0 and +100 km~s$^{-1}$.  
Emission features at negative velocities in Figure 2b are contributed by 
superposed 809~GHz $^3P_2 \rightarrow ^3P_1$ [C\,I] emission from the 
upper sideband (see \S 2).

For comparison, longitude-velocity maps of the $^{12}$CO $J = 1 \rightarrow 0$, 
$^{13}$CO $J =1 \rightarrow 0$, and CS $J = 2 \rightarrow 1$ data
described by Stark et al. (1987) and Bally et al. (1987, 1988) 
are shown in Figure 3.  Again, it is seen that the low-$J$ 
rotational transitions are broadly and similarly distributed, while the 
CO $J =7 \rightarrow 6$ emission is much more spatially confined.
A direct superposition of the $^{12}$CO $J = 4 \rightarrow 3$ and 
$J = 7 \rightarrow 6$ maps is shown in Figure 4, along with sample 
spectra at several positions. 

Figure 5 shows maps of $^{12}$CO $J = 7 \rightarrow 6$ to $^{12}$CO 
$J = 4 \rightarrow 3$ ($T^{12}_{7\rightarrow6}/T^{12}_{4\rightarrow3}$) 
and $^{12}$CO $J = 1 \rightarrow 0$ to $^{13}$CO $J = 1 \rightarrow 0$ 
($T^{12}_{1\rightarrow0}/T^{13}_{1\rightarrow0}$) line brightness temperature 
ratios as a function of longitude and velocity in the Galactic Center region. 
The data were all convolved to the resolution of the $^{12}$CO $J = 1 \rightarrow 
0$ map.
The data in this figure fall into three categories: (1) regions where the 
signal-to-noise ratio is high for both the numerator and the denominator, 
(2) regions where the signal-to-noise ratio is low for the numerator but 
high for the denominator, and (3) regions where the signal-to-noise ratio 
is low for both the numerator and denominator. The fourth possibility, 
where the signal-to-noise ratio is high for the numerator but low for the 
denominator, is avoided by placing the higher signal-to-noise data in 
the denominator. Category 1 data appear as smooth blue, green, yellow, 
orange, or white areas of the map. In these regions, the temperature ratio 
is well-determined, and has a fractional error only a little larger than 
that of the temperature data constituting the numerator. Category 2 data 
appear as mottled black and blue, since the value of the numerator is small 
and dominated by noise, while the denominator is some well-determined number 
([zero $\pm$ noise]/[number]); the values of the ratio in these regions 
are near zero, with large fractional error. 
Category 3 data have a ``salt and pepper'' black and white appearance as the ratio 
varies wildly over all possible values, both positive and negative, since both the
numerator and denominator have noisy values near zero. Both the value of the ratio 
and its error are large. In the upper panel of Figure 5, category 1 data can be found
in the vicinity of Sgr A and Sgr B; category 2 data is found between Sgr A and Sgr B  
and to the left of Sgr B; category 3 data occupies the periphery of the figure. 

Typical values of $T^{12}_{7\rightarrow6}/T^{12}_{4\rightarrow3} \approx\, 
0.3 \pm 0.05$ at points near Sgr~A.  
At positions on the strip near Sgr~B, typical values are higher:
$T^{12}_{7\rightarrow6}/T^{12}_{4\rightarrow3} \approx\, 0.6 \pm 
0.05$, over the velocity range 50 to 85 \kms\ . Average line temperature 
ratios near Sgr~A and Sgr~B are summarized in Table 1. 
The value of $T^{12}_{7\rightarrow6}/T^{12}_{4\rightarrow3} $ near Sgr~B is
a factor of two higher than the ratio observed in the nuclear region 
of the starburst galaxy M82 (Mao et al. 2000). 
Figure 5 shows that aside from the regions near Sgr~A and Sgr~B, 
$T^{12}_{7\rightarrow6}/T^{12}_{4\rightarrow3} \approx 0.12 \pm 0.08$ 
in the region bounded roughly by $0.1 < \ell < 1.5$ 
and $-$10~km~s$^{-1}$ $< v <$ 110~km~s$^{-1}$.
This is the area that shows significant CS $J = 2 \rightarrow 1$ emission 
in Figure 3 and is the area that Binney et al. (1991) identify with
$x_2$ orbits in the galactic bulge.  The 300 parsec ring is the rough 
parallelogram-shape between $\ell$=$-$1$^{\circ}$ and $\ell$=1.8$^{\circ}$, 
and velocities from $-$200 to $+$200 \kms . The Clump 2 molecular cloud 
complex is located near $\ell$=3$^{\circ}$, at velocities from $+$20 
to $+$150 \kms\ . Both these regions show 
$T^{12}_{7\rightarrow6}/T^{12}_{4\rightarrow3} < 0.1 $.

\subsection{LVG Analysis Using the Line Ratios}

Using the line ratios, we can estimate the kinetic temperature, $T_{kin}$, 
and the number density of molecular hydrogen, $n({\mathrm H_2})$, through 
a large velocity gradient (LVG) radiative transfer analysis 
(Goldreich \& Kwan 1974).  The LVG approximation simplifies radiative 
transfer by the assumption that an emitted spectral line photon can only 
be absorbed ``locally'', within a small region whose velocity is similar 
to the point of emission.  This approximation is robust, in the sense 
that the results of LVG models are often reasonably accurate, even when 
reality violates the assumptions underlying the models (Ossenkopf 1997). 
Our LVG radiative transfer code simulates a plane-parallel cloud geometry.
It uses the CO collisional rates from Turner (1995) and newly-derived values 
for the H$_2$ ortho-to-para ratio ($\approx$ 2) and the collisional quenching 
rate of CO by H$_2$ impact (Yan, Balakrishnan, \& Dalgarno, in preparation).
The model has two input parameters: the ratio of $^{12}$CO to $^{13}$CO 
abundance, and the ratio $X(CO) / \nabla V$ , where $X(CO)$ is the fractional 
CO abundance parameter and $\nabla V$ denotes the velocity gradient.
The abundance ratio $^{12}$CO/$^{13}$CO is taken to be 25 in the Galactic 
Center region (Langer \& Penzias 1990, 1993).  We take 
$X(CO) / \nabla V= 10^{-4.5} \, {\mathrm{pc \, km^{-1} \, s}}$, assuming 
that the $^{12}$CO/H$_2$ ratio is 10$^{-4}$ and the velocity gradient of 
Galactic Center clouds is typically 3 to 6 km s$^{-1}$pc$^{-1}$ (Dahmen 
et al. 1998). 

The line ratios for the LVG analysis and the physical parameters derived 
are summarized in Table 1. 
The solution of the radiative transfer equation gives 
$T^{12}_{7\rightarrow6}/T^{12}_{4\rightarrow3}$ and 
$T^{12}_{1\rightarrow0}/T^{13}_{1\rightarrow0}$ as a function of the total 
molecular hydrogen volume density for different values of the gas kinetic 
temperature, as shown in Figure 6. For each observed point, we can invert 
these functions to determine the kinetic temperature and molecular hydrogen 
volume density corresponding to the observed line ratios.

\subsection{Temperature and Density Variations in the Galactic Center}

The LVG model shows the variation of kinetic temperature across the Galactic 
Center region. Sgr~B, with $T^{12}_{7\rightarrow6}/T^{12}_{4\rightarrow3} 
\approx$ 0.6, has a kinetic temperature of 72 ($\pm$2) K, the highest 
kinetic temperature among the Galactic Center molecular clouds on our
strip map. This derived kinetic temperature is consistent with the 
rotational temperature derived from the NH$_3$(J,K)=(3,3) $\rightarrow$ 
(1,1) data (Morris 1989) and $^{13}$CO $J = 1 \rightarrow 0$ survey 
(Oka et al. 1998). The kinetic temperature at the position nearest Sgr~A 
is 47 ($\pm3$) K, lower than Sgr~B.  If we exclude these two regions, we 
find that the mean kinetic temperatures is $\sim$35 $\pm10$ K 
within the inner 500 parsecs of the Galaxy. In comparison, the dust 
temperatures in the inner 200 pc are $\sim$21$\pm$2 K (Pierce-Price 
et al. 2000). 

The observed line intensities of CO $J =1 \rightarrow 0$, $J =4 \rightarrow 
3$, and $J =7 \rightarrow 6$ lines, together with the LVG values for the 
kinetic temperature, allow us to estimate the column densities of all the 
$J$-states of CO.  The CO $J =4 \rightarrow 3$ line is the most luminous 
CO transition from the Sgr~A complex, whereas the $J=5 \rightarrow 4$ 
line is the most luminous CO transition from the Sgr~B complex. 

Adopting diameters of 45 pc for Sgr~B and 30 pc for Sgr~A from inspection 
of $^{13}$CO and CS maps (Bally et al. 1987), and assuming uniform 
thermodynamic properties within those volumes, we can estimate the 
properties of those clouds as a whole.  The total CO $J =7 \rightarrow 
6$ luminosity integrated over the Sgr~B complex is $\sim 5 \times 10 ^{36}$ 
ergs s$^{-1}$ and integrated over the Sgr~A complex is $\sim 1.5 \times 
10^{36}$ ergs s$^{-1}$.  We estimate the total CO rotational line cooling 
to be $L_{CO}  \approx  2.9 \times 10 ^{37}$ ergs s$^{-1}$ from 
Sgr~B and $L_{CO} \approx  2.0 \times 10^{37}$ ergs s$^{-1}$ from 
Sgr~A. The CO $J =7 \rightarrow 6$ transition accounts for 7.5\% 
of the total CO luminosity from the Sgr~A complex (Table 2). At the Sgr~B 
complex, with its higher gas temperature, the CO $J =7 \rightarrow 6$ 
transition contributes more than 17\% of the total CO luminosity.  The 
total estimated energy released by the CO rotational lines toward the 
Sgr~B cloud is, however, similar to that of Sgr~A.

Using the derived molecular hydrogen volume density from our LVG analysis 
and assuming spherical geometry, we can estimate the masses of the Sgr~B 
and Sgr~A molecular clouds. Multiplying the LVG-calculated density by the 
size of Sgr~B, we find its mass is approximately 10$^6$ ${\mathrm M}_\odot$, 
consistent with estimates of $\sim$8 $\times$ 10$^5$ ${\mathrm M}_\odot$, 
derived using dust continuum emission (Lis, Carlstrom, \& Keene 1991; 
Gordon et al. 1993). The CO $J = 7 \rightarrow 6$ emitting region towards 
the Sgr~A complex has a mass of $\sim 0.4 \times 10^6 {\mathrm M}_\odot$.

\section{Summary \& Conclusions}

We have presented maps of the Galactic Center region in the CO 
$J=7 \rightarrow 6$ and $J=4 \rightarrow 3$ transitions observed with 
the Antarctic Submillimeter Telescope and Remote Observatory (AST/RO) 
located at Amundsen-Scott South Pole Station. Comparing 
the CO $J=4 \rightarrow 3$, CO $J=1 \rightarrow 0$, and 
$^{13}$CO $J=1 \rightarrow 0$ maps reveals that the CO $J=7 \rightarrow 6$
emission is concentrated toward the Sgr~A and Sgr~B complexes. Using an 
LVG model, we find that the gas kinetic temperatures of the Sgr~B and Sgr~A 
complexes are 72 K and 47 K and the gas densities of those regions are 
10$^{4.15}$ cm$^{-3}$ and 10$^{3.9}$ cm$^{-3}$. Excluding those regions, 
we estimate the mean kinetic temperature and density throughout the 
Galactic Center are 30 to 45 K and 10$^{3.7\pm0.05}$ cm$^{-3}$. The CO 
$J =7 \rightarrow 6$ transition accounts for 7.5\% of the total CO 
luminosity from the Sgr~A complex. At the Sgr~B complex, with its higher 
gas temperature, the CO $J =7 \rightarrow 6$ transition contributes 
more than 17\% of the total CO luminosity.  
 
The tragic death of Rodney Marks, who was the year 2000 AST/RO Winterover 
Scientist, occurred just before this observational program.
The remaining winterover crew from the Center for Astrophysical Research
in Antarctica---Gene Davidson, Greg Griffin, David Pernic, and
John Yamasaki---continued AST/RO operations in tribute to
Rodney's memory, allowing these observations to be made.
The AST/RO group is grateful for the logistical support of the 
National Science Foundation (NSF), Antarctic Support Associates, 
Raytheon Polar Services Company, and the Center for Astrophysical 
Research in Antarctica during our polar expeditions. SK thanks Min Yan 
for LVG analysis and Wilfred Walsh for his helpful comments on the 
manuscript. This work was supported in part by  United States
National Science Foundation grant DPP88-18384, and by
the Center for Astrophysical Research in Antarctica and the NSF
under Cooperative Agreement OPP89-20223.

\begin{table*}
\caption{Parameters and Results of LVG Model Calculations}
\begin{tabular}{llll}
\noalign{\smallskip}
\hline\hline
\noalign{\smallskip}
Parameter          &~Sgr~A   &~Sgr~B \\
\noalign{\smallskip}
\hline
$T^{12}_{7\rightarrow6}/T^{12}_{4\rightarrow3}$ &$0.3\pm 0.05$ & $0.6\pm0.05$ \\
$T^{12}_{1\rightarrow0}/T^{13}_{1\rightarrow0}$ &$5.7\pm1.6$ &$ 5.0\pm1.0$  \\
Velocity Range (\kms\ )   & 10 $-$ 80  & 50 $-$ 85 \\
$n$(H$_2$) (cm$^{-3}$) &10$^{3.9\pm0.05}$ & 10$^{4.15\pm0.05}$ \\
$T_{kin}$ (K)     & $47\pm3$ & $72\pm2$ \\
Mass H$_2$ ($M_\odot$) & 0.4 $\times$ 10$^6$ & 1.0 $\times$10$^6$ \\  
Size (pc)        & 30                & 45     \\
\noalign{\smallskip}
\hline
\end{tabular}
\end{table*}

\begin{table*}
\caption{CO \& CI Luminosity}
\begin{tabular}{llll}
\noalign{\smallskip}
\hline\hline
\noalign{\smallskip}
                                       &    Sgr~A              &        Sgr~B         \\
\noalign{\smallskip}
\hline
$L$(CO 7$\rightarrow$6) (ergs s$^{-1}$) & 1.50$\times$10$^{36}$ & 5.07$\times$10$^{36}$ \\
$L$(CO 4$\rightarrow$3) (ergs s$^{-1}$) & 4.73$\times$10$^{36}$ & 5.59$\times$10$^{36}$ \\
$L$(CO total)   (ergs s$^{-1}$)        & 2.00$\times$10$^{37}$ & 2.92$\times$10$^{37}$ \\
$L$(C I total) (ergs s$^{-1}$)        & 5.46$\times$10$^{35}$ & 8.96$\times$10$^{35}$ \\
\noalign{\smallskip}
\hline
\end{tabular}
\end{table*}



\begin{figure}
\figurenum{1}
\plotone{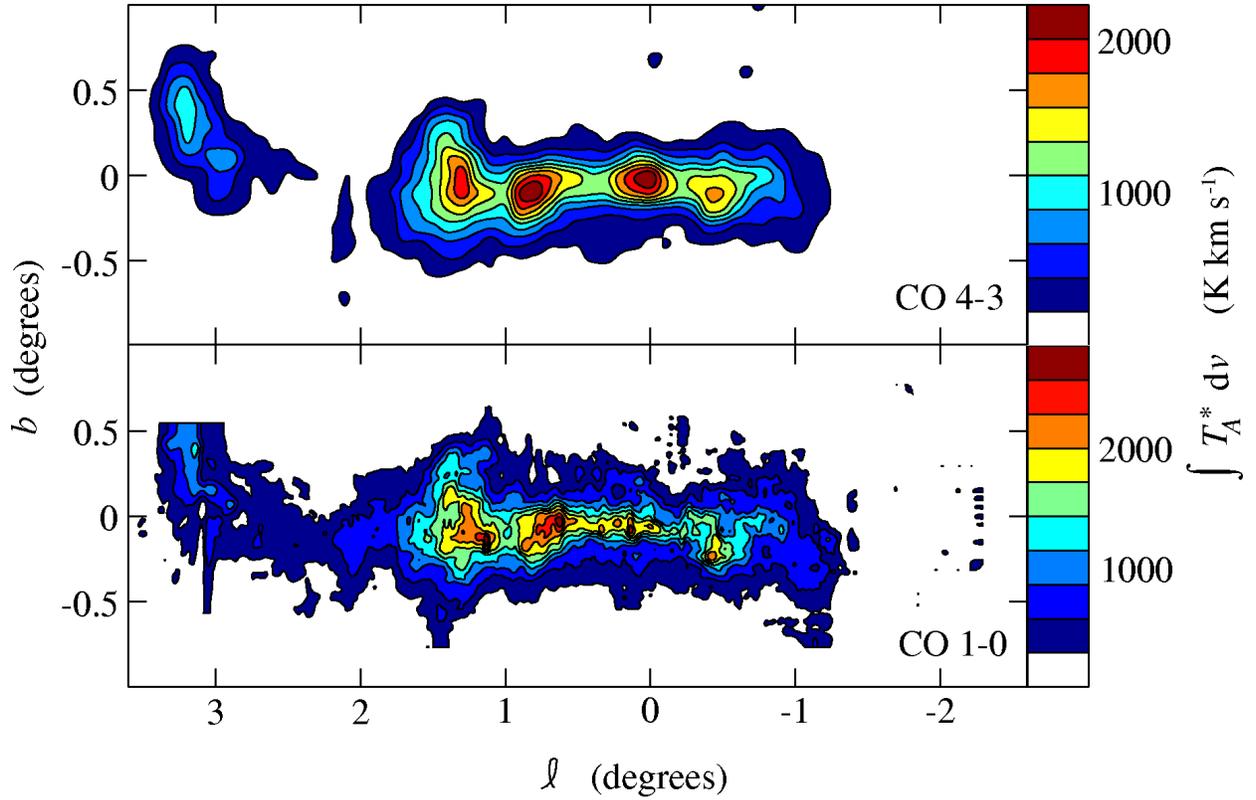}
\caption{{\it Top}: Integrated intensity map of the $^{12}$CO $J = 4 \rightarrow 3$
emission observed with the AST/RO telescope (Martin et al., in preparation). 
{\it Bottom}: Integrated intensity map of the $^{12}$CO $J = 1 \rightarrow 0$ emission 
from the same region (Bell Labs 7m data, Stark et al. 1987). 
Ten equally spaced contours are drawn at the levels shown in the colorbar above.}
\label{f:fig1}
\end{figure}

\begin{figure}
\figurenum{2}
\epsscale{0.9}
\plotone{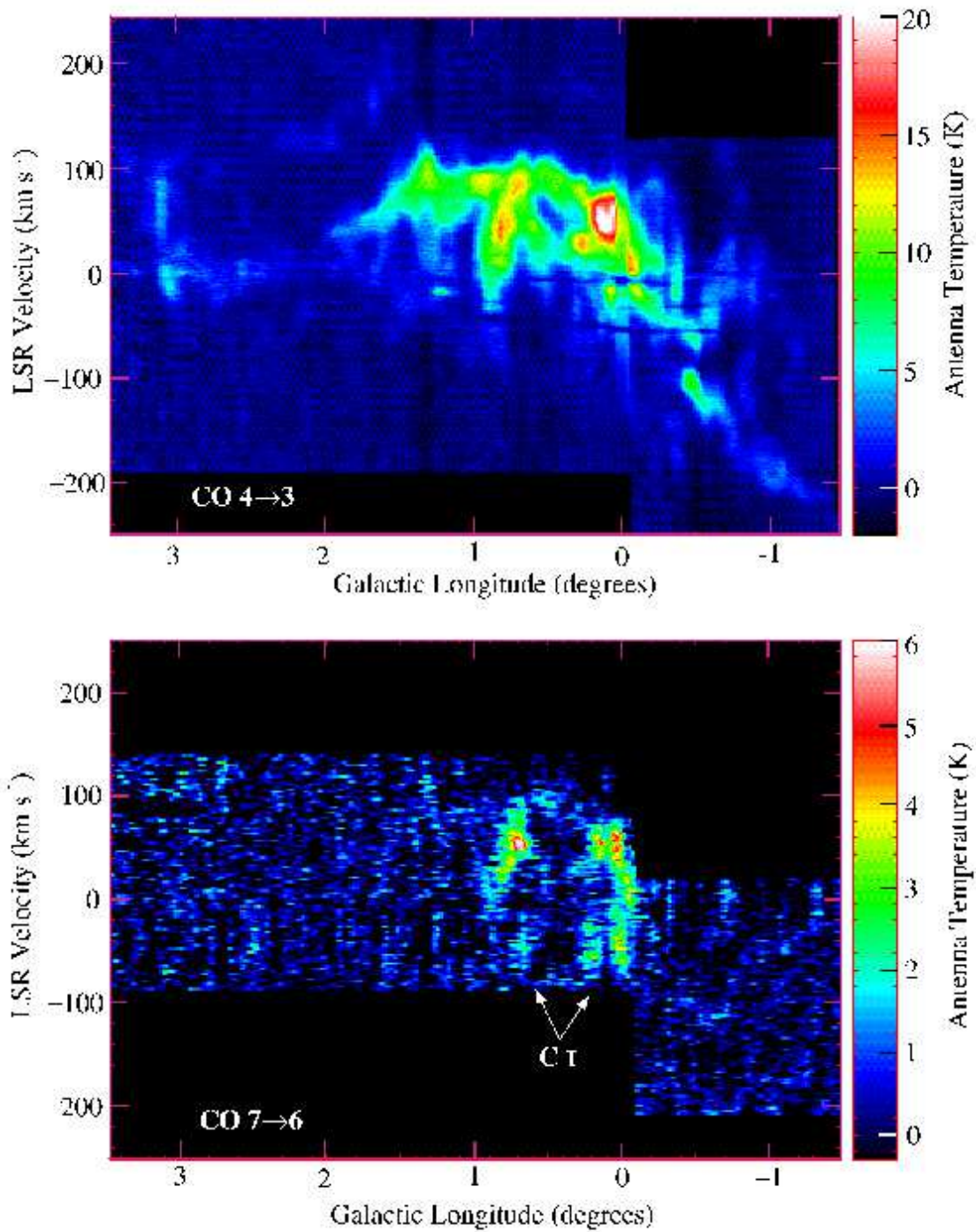}
\caption{Longitude-velocity maps of CO emission observed toward the Galactic
Center with AST/RO. {\it Top}: $J = 4 \rightarrow 3$. {\it Bottom}: $J = 7 
\rightarrow 6$. The emission components nominally appearing near $-50$ km s$^{-1}$
in the bottom map are from the 809 GHz [CI] line (see text).  The Sgr~A cloud is 
near $\ell = 0^{\circ}$ and the Sgr~B cloud is near $\ell = 0.67^{\circ}$.  The dark
horizontal bands in the top map near $v = 0$ to $-60$ km~s$^{-1}$ are due to absorption
in foreground gas.}
\label{f:fig2}
\end{figure}

\vspace*{-0.5cm}
\begin{figure}
\figurenum{3}
\epsscale{0.70}
\plotone{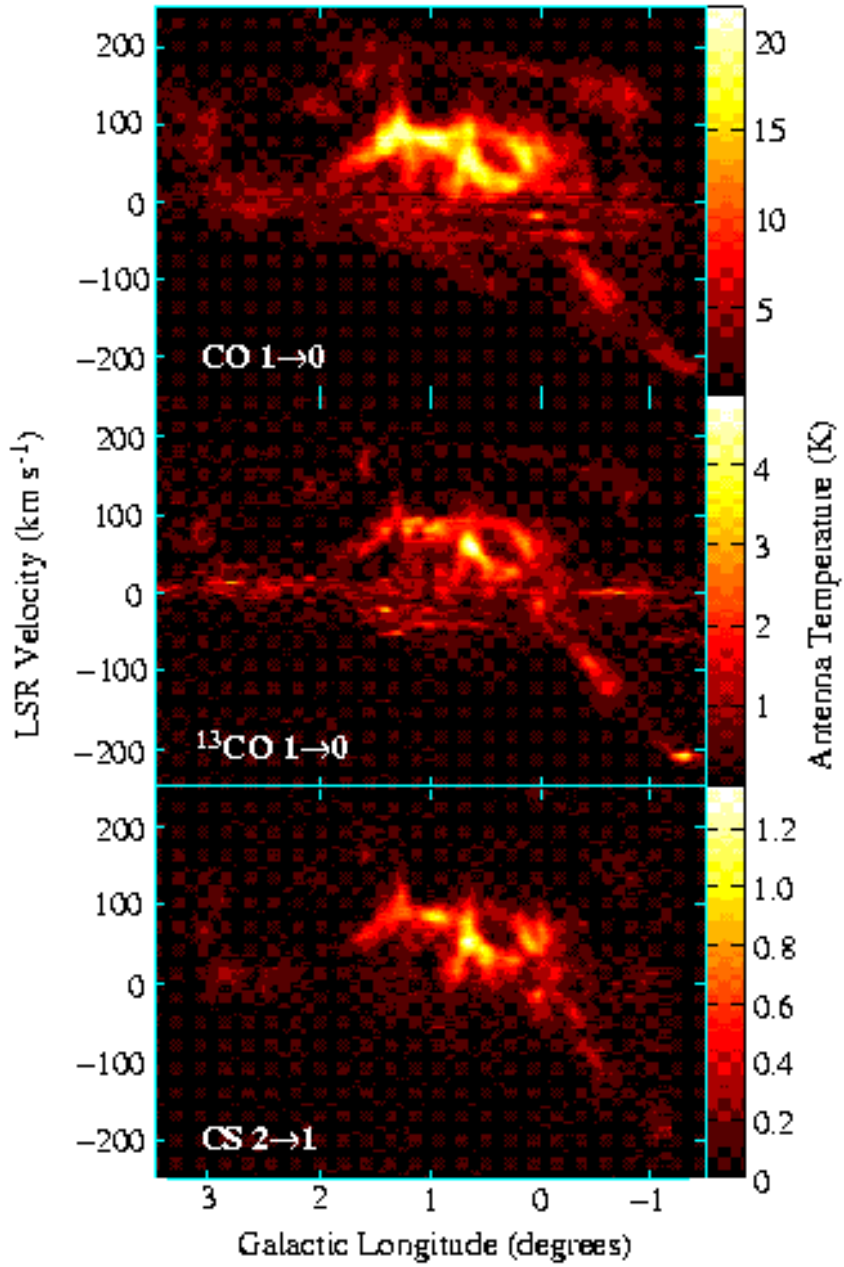}
\vspace*{-0.3cm}
\caption{Longitude-velocity maps of $^{12}$CO ($J = 1 \rightarrow 0$) emission 
({\em top}), from Bell Labs 7m data described by Stark et al. (1987); $^{13}$CO 
($J = 1\rightarrow 0$) ({\em middle}) and CS ($J = 2 \rightarrow 1$) emission 
({\em bottom}), from data presented by Bally et al. (1987). All three data 
sets are shown smoothed to $3'$ resolution.  The $^{13}$CO data for 
$\ell > 0.9^{\circ}$ and $\ell < -0.5^{\circ}$ were observed with a spacing of 
$6'$, leading to a coarse appearance in these regions.}
\label{f:fig3}
\end{figure}

\begin{figure}
\figurenum{4}
\epsscale{1.1}
\plotone{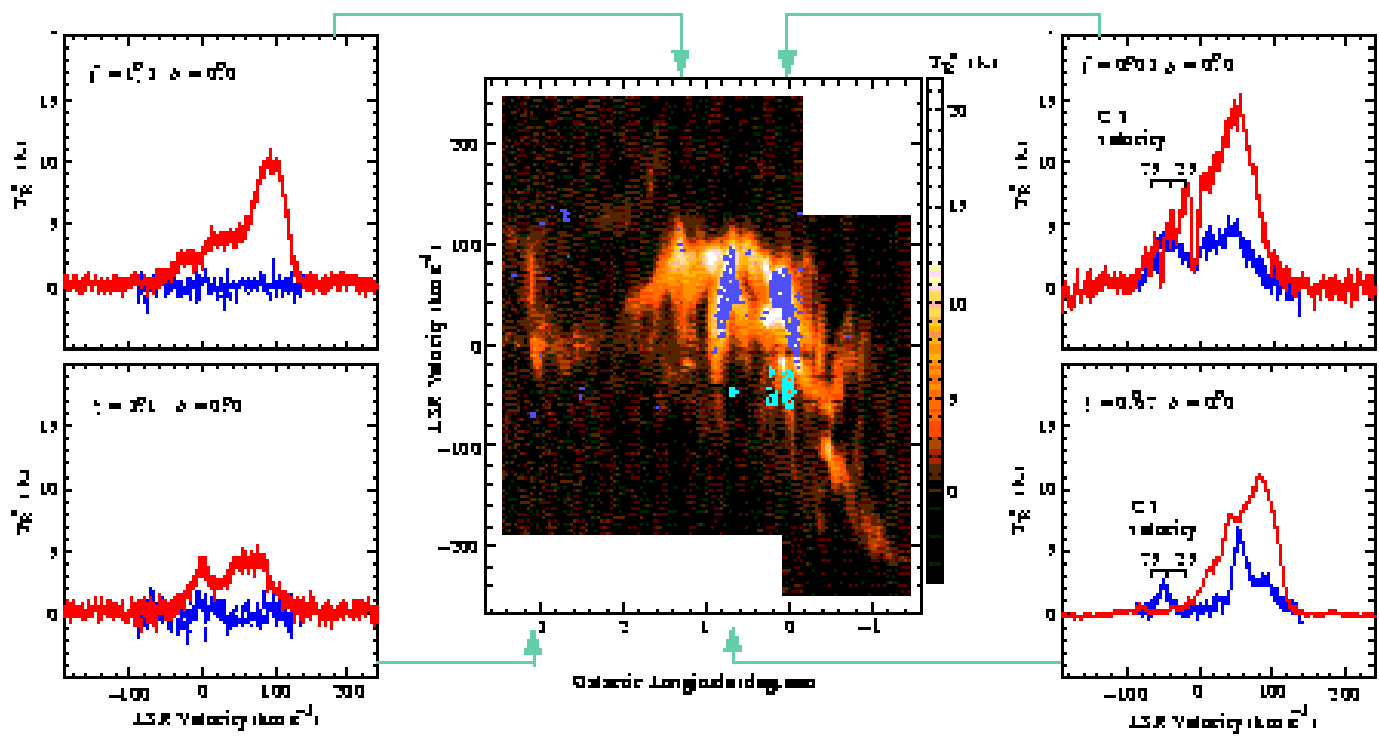}
\caption{Sample $^{12}$CO $J=7 \rightarrow 6$ ({\it blue lines}) and 
$^{12}$CO $J=4 \rightarrow 3$ ({\it red lines}) spectra observed toward 4 
different positions, as marked.  Center map shows the $^{12}$CO $J = 7 
\rightarrow 6$ contours (in blue) superposed on the
$^{12}$CO $J = 4 \rightarrow 3$ longitude-velocity map (in red). Superposed 
emission from the 809 GHz [CI] line in the upper sideband is shown in 
light blue.  The $^{12}$CO $J = 7 \rightarrow 6$ spectra are shown after 
Hanning smoothing. The spectra near Sgr~B ($\ell=0.67^{\circ}$) have higher 
signal-to-noise ratio due to significantly greater integration time than the
other, more typical spectra.}
\label{f:fig4}
\end{figure}

\begin{figure}
\figurenum{5}
\epsscale{0.9}
\plotone{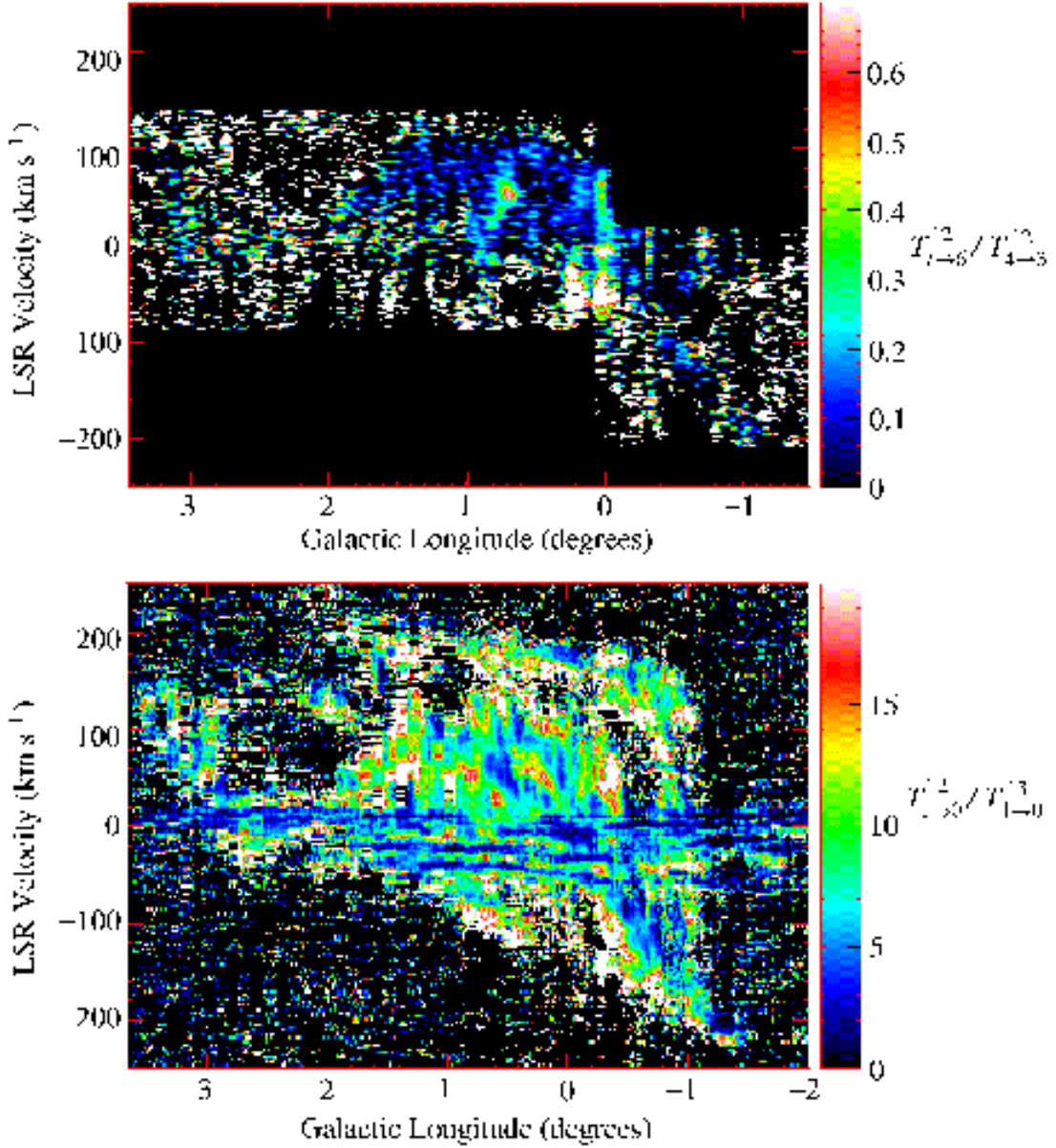}
\caption{Longitude-velocity maps of the ratios of line temperatures of CO 
transitions in the Galactic Center region:  
(a)  $T^{12}_{7\rightarrow6}/T^{12}_{4\rightarrow3}$ (this paper),
(b)  $T^{12}_{1\rightarrow0}/T^{13}_{1\rightarrow0}$ (data from Stark et 
al. 1987).} 
\label{f:fig5}
\end{figure}

\begin{figure}
\figurenum{6}
\epsscale{0.9}
\plotone{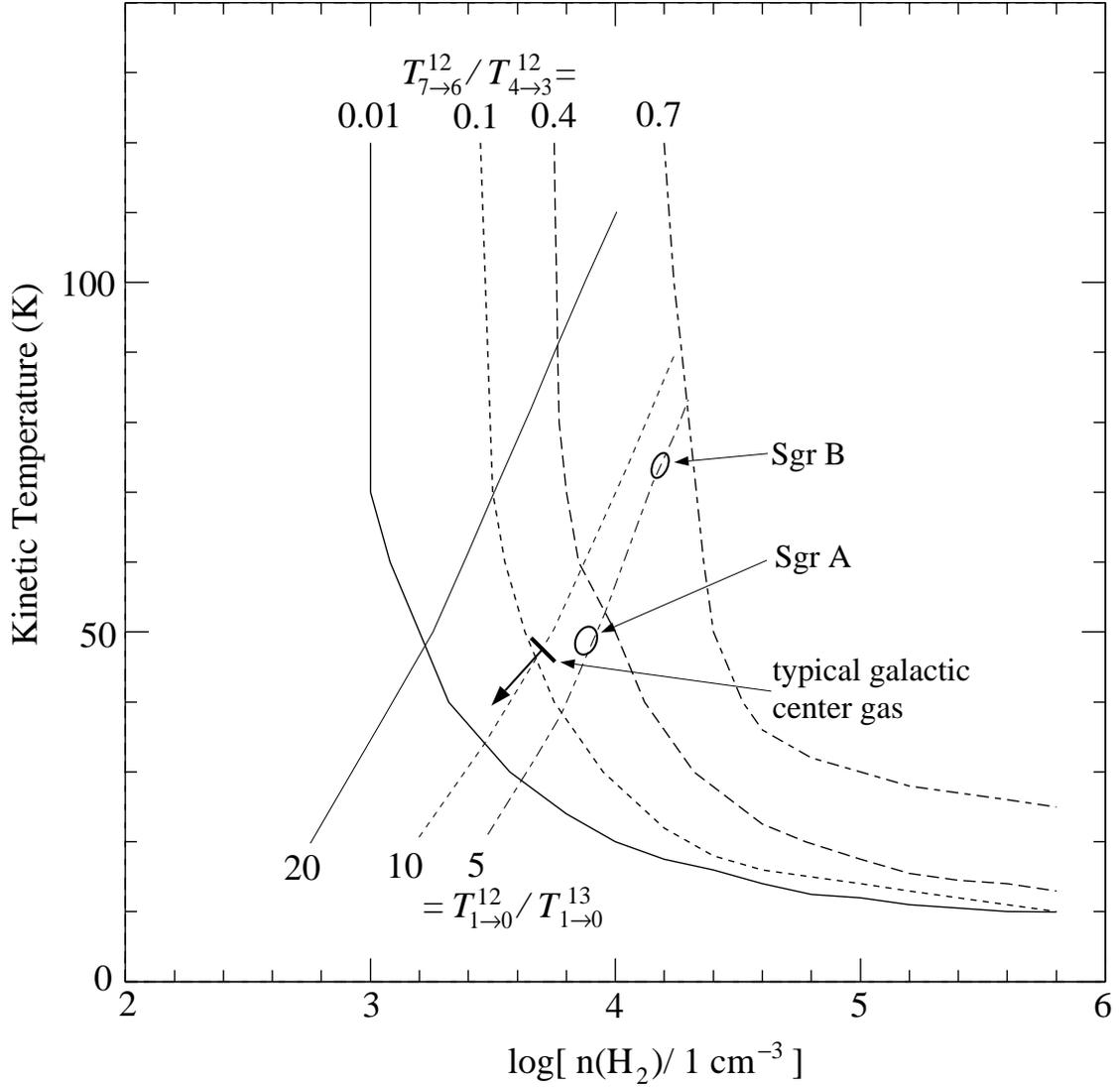}
\caption{Large Velocity Gradient (LVG) model results for 
$X(CO)/ \nabla V = 10^{-4.5} \mathrm{pc \, km^{-1} \, s}$ ($X(CO)$ is the 
fractional CO abundance parameter). Solid and dashed lines show constant 
values of $T^{12}_{7\rightarrow6}/T^{12}_{4\rightarrow3}$ and of 
$T^{12}_{1\rightarrow0}/T^{13}_{1\rightarrow0}$, as marked. 
Elliptical symbols represent the data with one sigma errors, from Table 1. 
The analysis uses newly derived values for the H$_2$ ortho-to-para ratio 
and for the collisional quenching rate of CO by H$_2$ impact 
(Yan, Balakrishnan, \& Dalgarno, in preparation).}
\label{f:fig6}
\end{figure}

\end{document}